# Cavity basics

*E. Jensen*
CERN, Geneva, Switzerland

**Abstract**
The fields in rectangular and circular waveguides are derived from Maxwell's equations by superposition of plane waves. Subsequently the results are applied to explain cavity modes. Interaction of the cavity modes with a charged particle beam leads to the fundamental parameters used to describe the performance of accelerating cavities. Finally an introduction to multi-gap cavities is given by the example of travelling-wave structures.

## 1   Introduction

The purpose of an RF cavity is to interact with charged particle beams in an accelerator. For the particles to gain energy, this interaction is the acceleration in the direction of particle motion, but special cavities exist to bunch, de-bunch, or re-bunch the beam, others to decelerate particles or to kick them sideways. For all these different applications, engineers are confronted with the task of optimizing the cavity for the given application according to certain design criteria within given constraints and without neglecting side-effects that might crucially deteriorate performance.

The design of RF cavities is a complex task involving understanding of beam physics, but also requires knowledge of the technologies used, design and construction methods, mechanics, materials, vacuum technology, high-voltage techniques, and many more. Sophisticated tools exist to calculate and design RF cavities, and this lecture on 'cavity basics' does not try to replace any of this. Its purpose is rather to explain the underlying basics and it is hoped that this will help the readers to develop a 'feeling' for the electromagnetic fields inside a cavity and the interaction with the particle beam.

In the following section we will start from Maxwell's equations to derive the electromagnetic fields in waveguides and subsequently demonstrate how we can obtain fields in cavities from them. The section can not apply the rigour that these derivations deserve in a full lecture, but it simplifies and illustrates wherever possible to allow the understanding of the basic concepts and ideas. In the following section we will introduce the main terms that are used in the description of cavities. Finally we devote the last section to an introduction to cavities with many gaps and in particular travelling-wave structures.

## 2   From plane waves to cavities

### 2.1   Homogeneous plane wave

The word 'cavity' is derived from the Latin *cavus* = hollow and describes, according to the Oxford English Dictionary, "A hollow place; a void or empty space within a solid body". Inside of the 'hollow place', the fields are governed by Maxwell's equations in free space:

$$\nabla \times \vec{B} - \frac{1}{c^2}\frac{\partial}{\partial t}\vec{E} = 0 \quad \nabla \cdot \vec{B} = 0 \quad ,$$
$$\nabla \times \vec{E} + \frac{\partial}{\partial t}\vec{B} = 0 \quad \nabla \cdot \vec{E} = 0 \quad . \tag{1}$$

We are ignoring the effect of the beam charge and current on the fields at this point. The curl of the third equation reads

$$\nabla \times \nabla \times \vec{E} + \nabla \times \frac{\partial}{\partial t}\vec{B} = 0, \tag{2}$$

while the time derivative of the first equation is

$$\frac{\partial}{\partial t}\nabla \times \vec{B} - \frac{1}{c^2}\frac{\partial^2}{\partial t^2}\vec{E} = 0. \tag{3}$$

Exchanging time and space derivative for the expression $\frac{\partial}{\partial t}\nabla \times \vec{B}$, Eqs. (2) and (3) can be combined to give

$$\nabla \times \nabla \times \vec{E} - \frac{1}{c^2}\frac{\partial^2}{\partial t^2}\vec{E} = 0. \tag{4}$$

Using the vector algebra identity $\nabla \times \nabla \times \vec{E} \equiv \nabla\nabla \cdot \vec{E} - \Delta\vec{E}$ and the second Maxwell equation we finally obtain the wave equation (or 4-dimensional Laplace equation):

$$\Delta\vec{E} - \frac{1}{c^2}\frac{\partial^2}{\partial t^2}\vec{E} = 0. \tag{5}$$

Solutions of this equation are homogeneous plane waves, i.e., solutions with separate trigonometric (or exponential) functions of the four independent variables space and time. A homogeneous plane wave oscillating at frequency $\omega$ and propagating in the space direction $\vec{k}$ can be written:

$$\begin{aligned}\vec{E} &\propto \vec{u}_y \cos(\omega t - \vec{k}\cdot\vec{r}), \\ \vec{B} &\propto \vec{u}_x \cos(\omega t - \vec{k}\cdot\vec{r}).\end{aligned} \tag{6}$$

One can easily verify that these expressions do in fact satisfy Eq. (5) if the constant resulting from the second time derivative $(\omega/c)^2$ is equal to the squared length of the vector $\vec{k}$, $k^2 = (\omega/c)^2$.

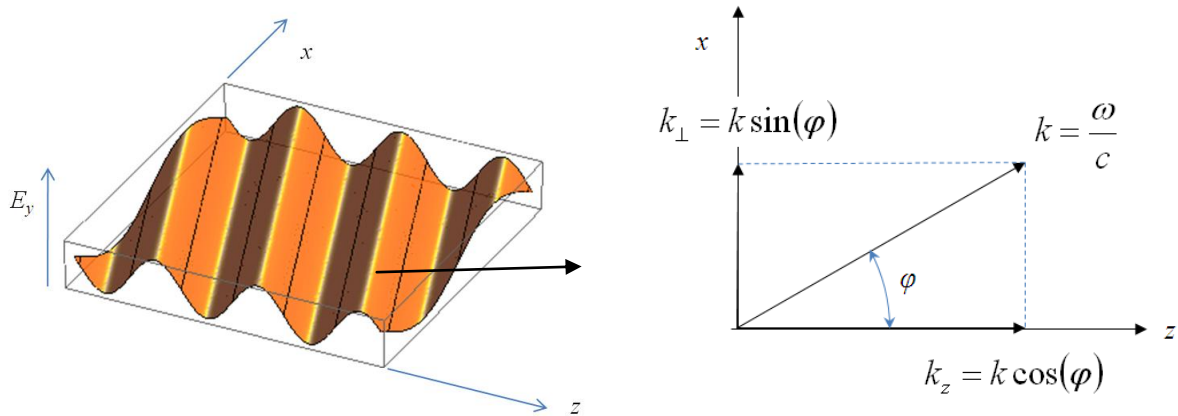

**Fig. 1:** A homogeneous plane wave, polarized in the *y* direction and propagating in the *x*–*z* plane. Left: snapshot of the wave travelling in the direction of the arrow; right: components of the wave vector

Interpreting the result, the vector $\vec{k}$ points in the direction of propagation of the plane wave oscillating at frequency $\omega$, its length $k$ measures the phase shift per unit length in this direction. Since a phase shift of $2\pi$ occurs over the free-space wavelength, we get the known relation

$$k = \frac{2\pi}{\lambda}. \tag{7}$$

$\vec{k}$ is known as the wave vector. If the wave propagates at an angle $\varphi$ with respect to a direction that we define to be the axial direction $z$, (see Fig. 1), we notice that the component of the wave vector in the $z$ and $x$ direction become $k_z = k\cos(\varphi)$ and $k_\perp = k\sin(\varphi)$, respectively. We have arbitrarily chosen the $y$ direction to be the polarization of the field (the direction of $\vec{E}$), but in order not to limit the general validity we use the symbol $\perp$ for the transverse direction, which we understand is orthogonal to both the electric field and the axial direction. We also note that since the components of the wave vector $k_\perp$ and $k_z$ are never larger than its length $k$, the wavelengths measured in those directions are always larger than or at least equal to the free-space wavelength.

The resulting relationship between the components of the wave vector (see the right-hand side of Fig. 1) is

$$k^2 = (\omega/c)^2 = k_\perp^2 + k_z^2, \tag{8}$$

a relationship which remains true for all hollow waveguides and allows one to calculate at a given frequency the propagation constant in the axial direction once the transverse components of the wave vector are known.

We have also marked the zero-crossings of the electric field in the left plot of Fig. 1 with thin black lines; these describe the (plane) phase fronts of the wave; with the oscillation in time with frequency $\omega$, these fronts travel with a speed $c$ in the direction of $\vec{k}$. The intersection of the phase front with the $z$ axis travels with a speed given by

$$v_{\varphi,z} = \frac{\omega}{k_z}, \tag{9}$$

which is always larger than or at least (in the limiting case $\varphi = 0$) equal to the speed of light.

RF cavities are linear and time-invariant (LTI) systems. The time invariance allows separating the time dependence out of Maxwell's equations; linearity allows the application of linear superposition. We will apply superposition of waves in the following, but also superposition of oscillations at different frequencies is possible, which allows application of the very powerful concept of Fourier transforms. This in turn allows describing problems and their solutions alternatively in the frequency domain or in the time domain, whichever is more suitable for a given problem.

Since we are now looking at oscillations at a single frequency, we do not need to carry the time dependence through the calculations at all — this is equivalent to looking at the Fourier component at frequency $\omega$. Furthermore, the oscillation at a fixed frequency is most conveniently described as a rotation in the complex plane, $\underline{a}\,e^{j\omega t}$, with a generally complex amplitude $\underline{a}$. Complex algebra allows now all calculations with just complex amplitudes; the 'real' time-dependent quantity is then obtained at the end by multiplying the complex amplitudes with $e^{j\omega t}$ and taking the real part. Phase relationships between different signals are then automatically taken care of. Another particular convenience of this approach is that the operators describing the time derivative and integral become simple multipliers:

$$\frac{\partial}{\partial t}(\ ) \equiv j\omega \cdot (\ ), \tag{10}$$

$$\int (\ )\,dt \equiv \frac{(\ )}{j\omega}. \tag{11}$$

It should further be noted that, when using this complex notation, the propagation of a wave in a direction $z$ simply now is described by the term

$$e^{-jk_z z}. \tag{12}$$

This will become useful when integrating the field along a particle trajectory with the correct consideration of the finite speed of the particle (see Section 3.3 below).

## 2.2 Superposition of two homogeneous plane waves

Now consider two plane waves, both polarized in the $y$ direction as in the example above, propagating in the $x$–$z$ plane, one at an angle $+\varphi$, the other at an angle $-\varphi$ with respect to the axial direction $z$, as sketched on the left of Fig. 2. The superposition of the two waves is sketched on the right-hand side;

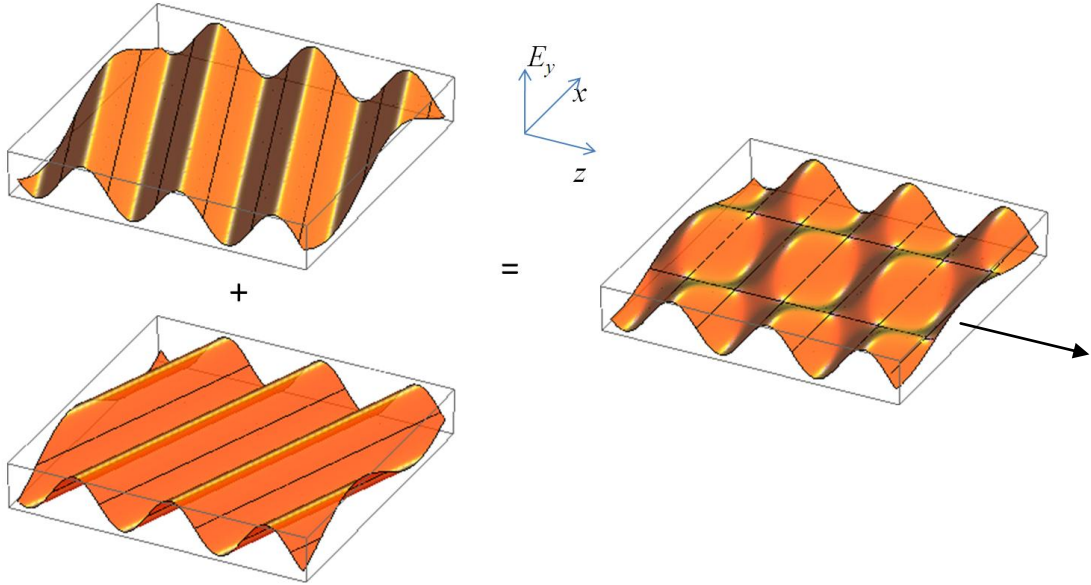

**Fig. 2:** The superposition of two plane waves. Left: two plane waves propagating at different angles; right: the superposition of these waves

note that the superposition of the two plane waves results in a wave pattern that is now travelling in the $z$ direction, while it forms a standing-wave pattern in the transverse $x$ direction. It is not visible in the snapshot of Fig. 2, but can be demonstrated in the equations. The field of a single plane wave travelling in direction $+\varphi$ is

$$E_{y+} \propto \sin(\omega t - k_z z + k_\perp x), \tag{13}$$

the one travelling at $-\varphi$ is given by

$$E_{y-} \propto \sin(\omega t - k_z z - k_\perp x). \tag{14}$$

The superposition of these two terms, using the trigonometric addition formula, is

$$E_{y+} + E_{y-} = 2\sin(\omega t - k_z z)\cos(k_\perp x). \tag{15}$$

As can be clearly seen from Eq. (15), the superposition has a standing-wave pattern in the transverse direction *x*, while the wave is travelling in the *z* direction with a phase velocity given by Eq. (9).

### 2.3 From superposition of plane waves to waveguide modes

We also see from the right-hand side of Fig. 2 and from the last term of Eq. (15) that we now have special transverse positions *x* where the electric field is always zero, i.e., the zeros of $\cos(k_\perp x) = 0$. At these positions *x* (the closest ones to the centre are at $x = \pm \dfrac{\pi}{2k_\perp}$) we might introduce a perfect conducting wall in the *y–z* plane without disturbing the field distribution. Figure 3 demonstrates this fact, showing the same field distribution as in Fig. 2, but with those boundaries inserted. With the distance *a* between these two boundaries, the transverse component can now only have discrete values, which are the integer multiples of $\pi/a$.

With our chosen *y* polarization for the electric field, we can also put perfectly conducting walls at any location $y = \text{const.}$, i.e., in the *x–z* plane. We have thus found valid, non-vanishing solutions of Maxwell's equations inside a **rectangular waveguide**. Since there are many solutions to the equation $\cos(k_\perp x) = 0$, there are many possible positions to put metallic walls in the transverse plane — they will have a different number of half-waves in the transverse direction. These different solutions are different **waveguide modes**.

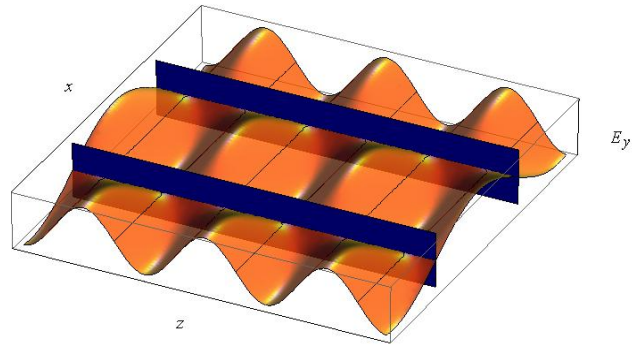

**Fig. 3:** Perfectly conducting boundaries can be inserted where the electric field is always zero; thus we have created a rectangular waveguide

The above example chosen for illustration was a special case with a certain polarization of the electric field and using the superposition of just two homogeneous plane waves propagating in the *x–z* plane; the modes we obtained are referred to as TE$_{n0}$-modes, where *n* denotes the number of half-waves in the *x* direction. The TE$_{10}$ mode of a rectangular waveguide has special importance since it has the lowest cutoff frequency; it is generally referred to as the fundamental mode.

More general solutions can be found using other polarizations and/or the superposition of more than two plane waves. Also the choice of the angle *φ* to remain in the *x–z* plane was arbitrary. The principle, however, that waves in rectangular waveguides can be constructed by superposition of homogeneous plane waves can be generalized — the result of this generalization is sketched in Fig. 4; it shows the possible discrete values for transverse wave vectors as points; the propagation constant can now be determined graphically for any frequency using Eq. (8), which now, for the mode with indices *m* and *n* in the *x* and *y* direction, respectively, becomes

$$k^2 = \left(\frac{\omega}{c}\right)^2 = k_\perp^2 + k_z^2 = \left(\frac{m\pi}{a}\right)^2 + \left(\frac{n\pi}{b}\right)^2 + k_z^2. \tag{16}$$

In Eq. (16), $a$ and $b$ denote the width and height of the rectangular waveguide, respectively. The points corresponding to the example illustrated above (the TE$_{n0}$-modes) lie on the abscissa of this plot. The TE$_{10}$ mode is the one closest to the origin, where $|k_x| = \pi/a$ and $k_y = 0$. For an arbitrary combination $m$ and $n$, the points inside a circle of radius $k$ will result in a positive $k_z^2$ (propagating mode), the ones outside result in a negative $k_z^2$ (evanescent mode). Note that for frequencies $f < c/2a$ (the lowest cutoff frequency), all modes are below cutoff (evanescent). Below this lowest cutoff frequency, waves cannot propagate in the waveguide.

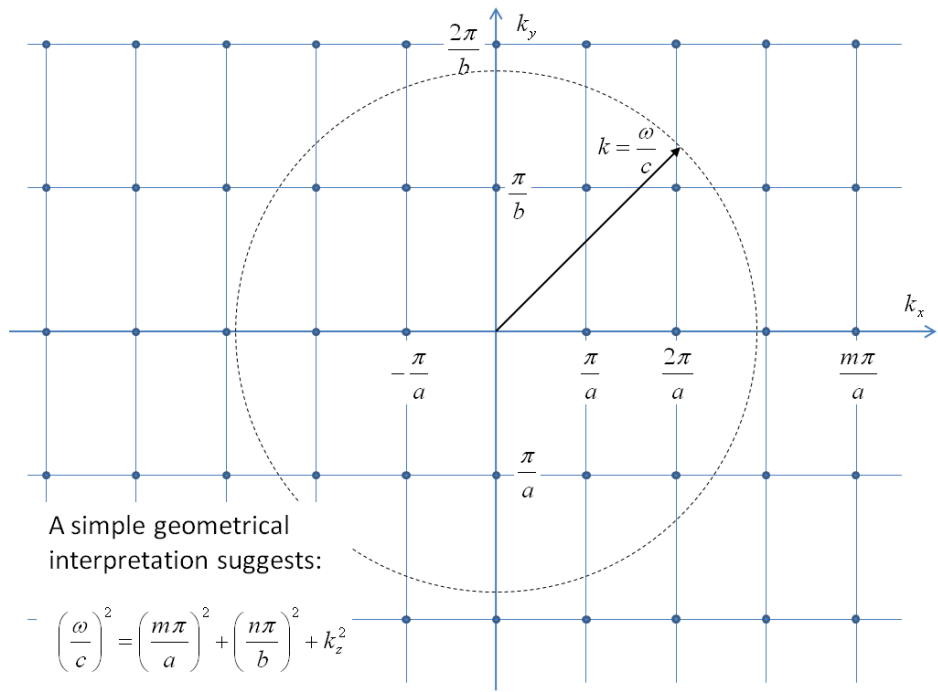

A simple geometrical interpretation suggests:

$$\left(\frac{\omega}{c}\right)^2 = \left(\frac{m\pi}{a}\right)^2 + \left(\frac{n\pi}{b}\right)^2 + k_z^2$$

**Fig. 4:** Cut-off wave numbers can be represented as points in the $k_x$–$k_y$ plane. Points inside the circle of radius $k = \omega/c$ represent modes that can propagate, points outside are not propagating or evanescent

The distribution of the electric field in the transverse plane of the first 12 modes (sorted by their cutoff frequency for an aspect ratio $a/b = 2$) in a rectangular waveguide is plotted in Fig. 5. Note that for TM modes the smallest possible index is 1.

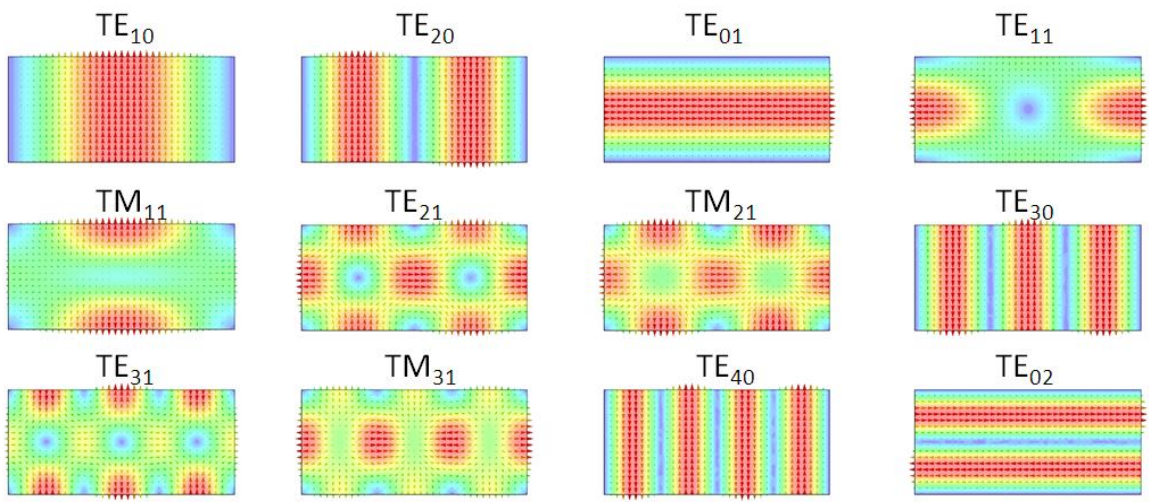

**Fig. 5:** The first 12 modes in a rectangular waveguide. The arrows of the electric fields are sketched

## 2.4 Circular waveguides

The idea of superposition of two homogeneous plane waves can be extended to the superposition of infinitely many plane waves to get other field configurations, e.g., those of a circular waveguide. To construct the $TM_{01}$ mode of a circular waveguide, for example, imagine again a plane wave as sketched in Fig. 1 above. Now consider $y$ as the axial, $x$ and $z$ as transverse coordinates. Equivalent to Eq. (14), we write

$$E_y(\varphi) \propto \sin(\omega t - k(z\cos(\varphi) + x\sin(\varphi))) \tag{17}$$

for a wave travelling in direction $\varphi$ with respect to the $z$ direction. If we now express also the transverse coordinates $z$ and $x$ in cylindrical coordinates $z = \rho\cos(\vartheta), x = \rho\sin(\vartheta)$, we get

$$E_y(\varphi) \propto \sin(\omega t - k\rho(\cos(\vartheta)\cos(\varphi) + \sin(\vartheta)\sin(\varphi))). \tag{18}$$

If we now add (integrate) the contributions of plane waves propagating in all directions $\varphi$, we get

$$\frac{1}{2\pi}\int_0^{2\pi} E_y(\varphi)\mathrm{d}\varphi \propto \frac{1}{2\pi}\int_0^{2\pi} \sin(\omega t - k\rho\cos(\varphi - \vartheta))\mathrm{d}(\varphi - \vartheta). \tag{19}$$

This latter integral is well known — it is defining the Bessel function of order zero:

$$\frac{1}{2\pi}\int_0^{2\pi} \sin(\omega t - k\rho\cos(\phi - \vartheta))\mathrm{d}(\phi - \vartheta) = J_0(k\rho)\sin(\omega t). \tag{20}$$

Similar to Fig. 3 above, the result of this superposition is illustrated in Fig. 6. One now obtains a radial standing-wave pattern, described by the Bessel function $J_0(k\rho)$. Just like the trigonometric functions, $J_0(k\rho)$ has zeros, so there exist radial positions where the electric field is always zero. At these radii, perfectly conducting boundary conditions may again be inserted without perturbing this field distribution — we have constructed a **circular waveguide**!

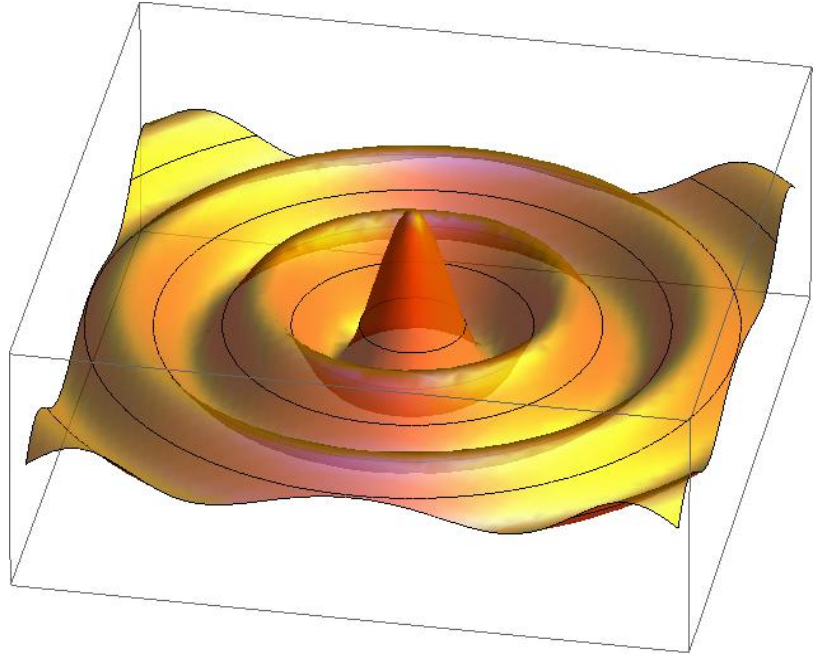

**Fig. 6:** Bessel function $J_0(k\rho)$, which can be obtained by superimposing homogeneous plane waves. Metallic walls may be inserted where $J_0(k\rho) = 0$

The superposition of plane waves was illustrated here for the special case of a wave polarized in axial direction and propagating in the transverse plane. This is — as was the example of the two waves in the rectangular waveguide above — a special case for the circular waveguide. Like for the rectangular waveguide many modes exist in round waveguides; they also are of the transverse electric (TE) and transverse magnetic (TM) type with respect to the axis, and they are also indexed with two numbers: the first for the azimuthal, the second for the radial 'number of half-waves'. The difference is that the cutoff wave numbers in the case of a rectangular waveguide are integer multiples of a constant (since the zeros of the trigonometric or harmonic functions are harmonics), while the zeros of the Bessel function have non-harmonic ratios. The first five zeros of $J_0(x)$, for example, are 2.40483, 5.52008, 8.65373, 11.7915, 14.9309, ...; they determine the cutoff frequencies of the $TM_{0n}$ modes. For large $n$, the zeros can be approximated by $\chi_{0n} \approx (n - 1/4)\pi$, so they become again 'almost harmonic'. The cutoff wave numbers of TE modes in circular waveguides are calculated from the zeros not of the Bessel function itself, but of its derivative.

The first of the zeros of the Bessel function and its derivative are marked in Fig. 7. Note that the lowest order mode in a circular waveguide is the $TE_{11}$ mode.

The equivalent of Eq. (16) in the case of circular waveguides is

$$k^2 = k_\perp^2 + k_z^2 = \left(\frac{\chi_{mn}}{a}\right)^2 + k_z^2 \quad . \tag{21}$$

where $\chi_{mn}$ is the $n$-th zero of the Bessel function of order $m$ in case of TM modes, or the $n$-th zero of the derivative of the Bessel function of order $m$ in case of TE modes, $a$ is the radius of the circular waveguide.

The equivalent to the dashed circle in Fig. 4 now is the dashed vertical line in Fig. 7, representing a given value for $ka$; the dots to the left of the vertical line are modes that can propagate, while the others require a negative $k_z^2$ to satisfy Eq. (21), which leads to non-propagating, evanescent modes.

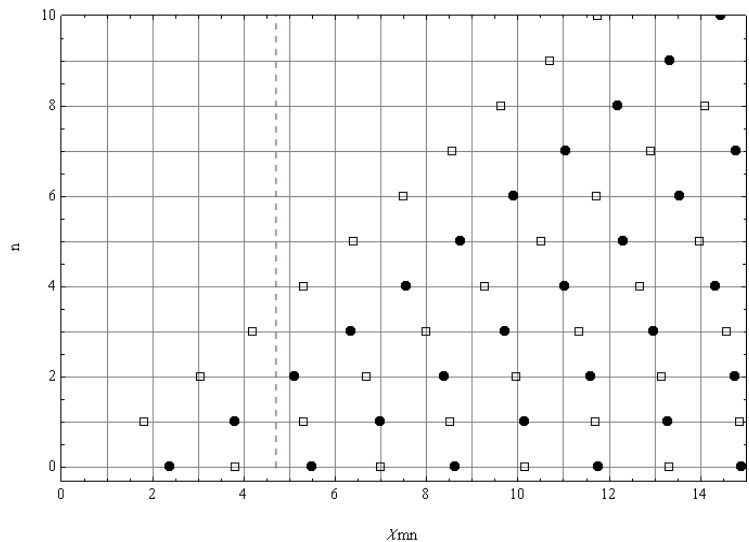

**Fig. 7:** Cutoff wave numbers for modes in a round waveguide. Closed circles: TM-modes; open squares: TE-modes

The field distributions (transverse electric field) of the six lowest order modes of a circular waveguide, again sorted by their cutoff frequencies, are plotted in Fig. 8. Please note that all modes with an azimuthal index $m \neq 0$ appear twice; this corresponds to the two independent transverse polarizations. Such a pair of modes corresponds to a double eigenvalue of the transverse wave equation, the two modes are said to be 'degenerate'. Any polarization (linear, circular or generally elliptical) can be constructed as superposition of the two shown polarizations with different amplitudes and phases — any of those combinations have the same eigenvalue and the same mode index.

Of particular interest amongst those modes in circular waveguides for accelerating cavities is of course the $TM_{01}$ mode, since it is rotational symmetric and has an axial field on axis. A short piece of circular waveguide operated in this mode is in fact a simple form of an accelerating cavity as we will see below.

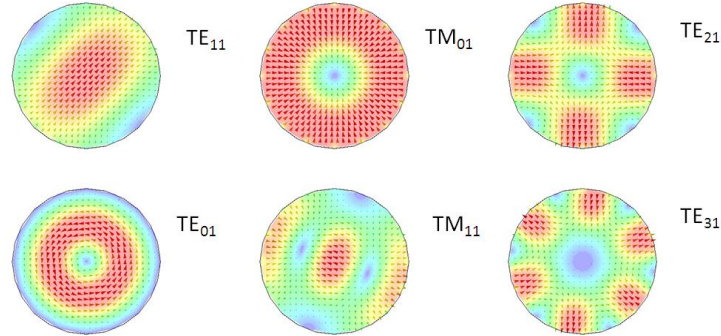

**Fig. 8:** The first six modes in a circular waveguide. The arrows of the electric fields are sketched. Modes with azimuthal index $m \neq 0$ appear with two different polarizations, only one of which is shown

Another set of modes of special interest, but this time for microwave power transport or for cavities optimized for storage of electromagnetic energy, are the $TE_{0n}$ modes, the first of which is the fourth mode shown above. These modes have no electric field at all at the metallic wall, which results in the interesting property that these modes have lower losses than any other mode. The $TE_{01}$ mode is used for low-loss power transport through waveguides over large distances. Note that the $TE_{01}$ mode is degenerate with the pair of $TM_{11}$ modes.

## 2.5 From waveguide to cavity

Starting from a round waveguide, and here in particular with the $TM_{01}$ mode, we can construct a cavity simply from a piece of waveguide at its cutoff frequency; this results in the fundamental mode of the so-called pillbox cavity, referred to as $TM_{010}$ mode. The electric and magnetic field of the $TM_{010}$ mode are plotted in Fig. 9. The eigenfrequency of this mode is the cutoff frequency of the waveguide and thus independent of the cavity height $h$. There is no axial field dependence, indicated by the axial index 0. The fields of the $TM_{010}$ mode in a simple pillbox cavity (closed at $z = 0$ and $z = h$ and at $r = a$ with perfectly conducting walls) are given by

$$E_z = \frac{1}{j\omega\varepsilon_0} \frac{\chi_{01}}{a} \sqrt{\frac{1}{\pi}} \frac{J_0\left(\frac{\chi_{01}}{a}\rho\right)}{a J_1\left(\frac{\chi_{01}}{a}\right)} ,$$

$$B_\phi = \mu_0 \sqrt{\frac{1}{\pi}} \frac{J_0\left(\frac{\chi_{01}}{a}\rho\right)}{a J_1\left(\frac{\chi_{01}}{a}\right)} ,$$

(22)

all other field components are zero. $\chi_{01}$ is the first zero of $J_0(x)$, $\chi_{01}$ = 2.40483. Note that the axial field, which is relevant for acceleration, is oscillating in phase at all positions *z*, corresponding to an infinite phase velocity in the axial direction. Note also that the electric and magnetic fields are out of phase by 90°, as indicated by the *j* in Eq. (22).

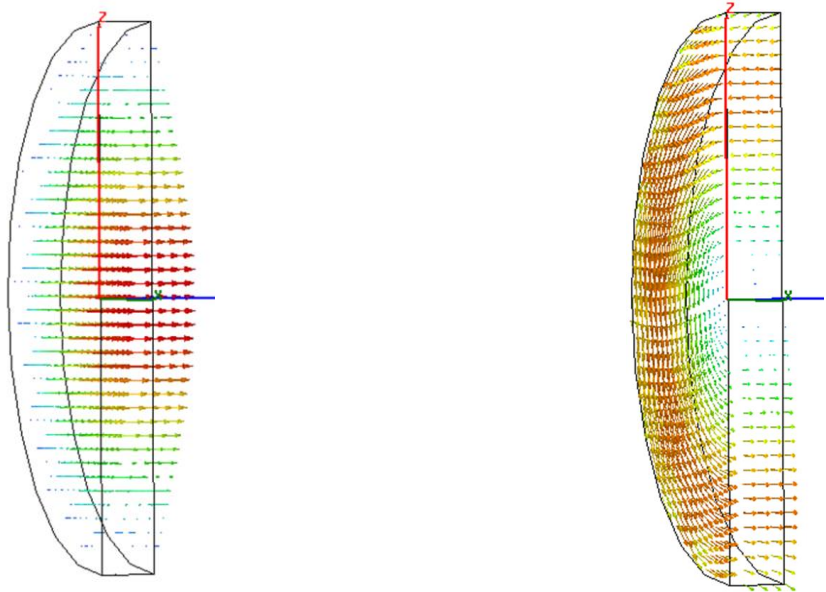

**Fig. 9:** A simple pillbox cavity, showing the electric field vector (left) and the magnetic field vector (right)

Other oscillation modes exist in a cavity formed by a piece of waveguide short-circuited at both sides; they are referred to as TM$_{mnp}$ and TE$_{mnp}$ modes; the two first indices indicate the waveguide mode as introduced above, the third index is the number of half-waves in the axial direction. Referring to Fig. 4, where we saw how the transverse boundaries allowed only discrete values for the transverse components of $\vec{k}$, the limitation also in the third direction allows only discrete values for the total vector $\vec{k}$, one might think of an extension of Fig. 4 to three dimensions. Discrete values of $\vec{k}$ of course mean discrete values of $\omega$ — these are the eigenmodes of the cavity.

The terms eigenmode and eigenfrequency stem from the fact that the mathematical problem to find non-vanishing solutions of Maxwell's equations without excitation is an eigenvalue problem. In the more general case of a closed cavity of arbitrary shape, which cannot be generated simply by a piece of waveguide, the determination of the oscillation modes is more difficult and generally done with the help of numerical simulation programs, but the general principle remains.

The cavities considered so far are entirely closed — this leaves no openings for the beam. Another possibility to create a cavity from a piece of waveguide is to provide perturbations in the waveguide cross section that lead to reflections of the waves; between two such perturbations, waves will be bouncing back and forth and thus equally form a cavity. As illustrated in Fig. 10, an oscillation mode forms between the two perturbations formed by the 'notches' in the cavity wall from the otherwise unperturbed TM$_{01}$ waveguide mode. The signal flow chart below demonstrates how this oscillation mode forms between the two perturbations. The shown oscillation mode would be called the TM$_{011}$ mode.

Only a fraction of the forward-travelling wave is reflected from the perturbation on the right in Fig. 10, and only a fraction of the backward-travelling wave is reflected from the perturbation on the left. This means that there is energy 'leaking' from the oscillation mode by diffraction into the pipe; this power loss will lead to a damped oscillation, subject of the next section. Note, however, that cavities between discontinuities in the beam chamber can form also involuntarily in an accelerator pipe; they are sometimes referred to as 'trapped modes' and may lead to beam instabilities. So also non-RF components installed in an accelerator need care, since they may deteriorate performance due to their possible cavity-like behaviour.

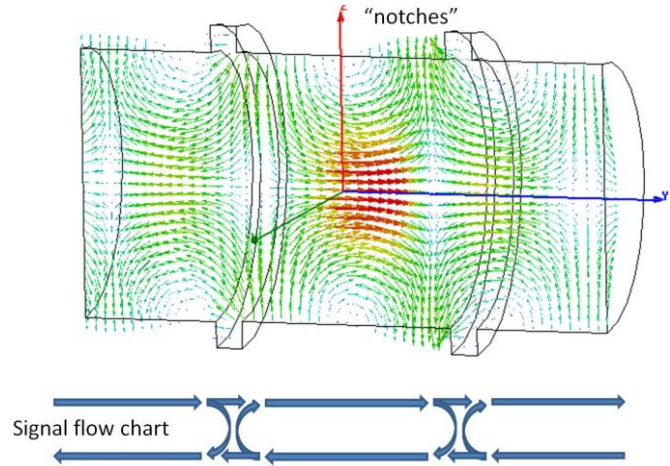

**Fig. 10:** A circular waveguide with a propagating TM$_{01}$ mode, perturbed by two notches

## 3 Characterizing a cavity

### 3.1 Stored energy

With the electric field energy $\iiint\limits_{cavity} \frac{\varepsilon}{2}|\vec{E}|^2 \, dV$ and the magnetic field energy $\iiint\limits_{cavity} \frac{\mu}{2}|\vec{H}|^2 \, dV$, the total energy stored is the sum of these two terms. As illustrated in Fig. 11, the energy is constantly swapping back and forth between these two energy forms at twice the RF frequency; while one is varying in time as $\sin^2(\omega t)$, the other one is varying as $\cos^2(\omega t)$, such that the sum is constant in time. Note also that this is related to the fact that $\vec{E}$ and $\vec{H}$ are exactly in quadrature, as we had already seen in Eq. (22) for the simple pillbox cavity.

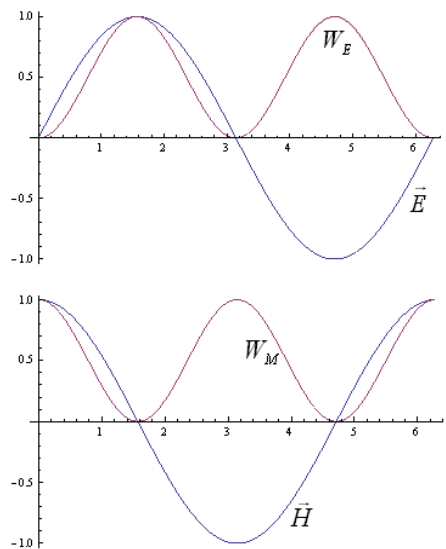

**Fig. 11:** Electric (top) and magnetic (bottom) stored energy in a cavity

## 3.2 Quality factor $Q$

As we said before, if the RF cavity would be entirely closed by a perfect conductor and the cavity volume would not contain any lossy material, there would exist solutions to Maxwell's equations with non-vanishing fields even without any excitation. These eigensolutions are the cavity (oscillation) modes introduced above. Each mode is characterized by its (eigen-)frequency and its characteristic field distribution inside the cavity.

If, however, the cavity walls are made of a good rather than a perfect conductor, or if the cavity is open as in Fig. **10**, modes still exist and are useful to characterize the cavity, but their eigenfrequencies will become complex, describing damped oscillations, so each mode will be characterized by its frequency and its decay rate. If the field amplitudes of a mode decay as $\propto e^{-\alpha t}$, the stored energy decays as $\propto e^{-2\alpha t}$. The quality factor $Q$ is defined as

$$Q = \frac{\omega_0 W}{P_{loss}} = \frac{\omega_0 W}{-\dfrac{d}{dt}W} = \frac{\omega_0}{2\alpha}. \tag{23}$$

Here $\omega_0$ denotes the eigenfrequency and $W$ the stored energy. $P_{loss}$ is the power lost into the cavity walls (or any other loss mechanism); it is equally the power that will have to be fed into the cavity in order to keep the stored energy at a constant value $W$. It is clear that the larger the $Q$, the smaller will become the power necessary to compensate for cavity losses. In other words, one can design the cavity to be operated at or near one of its eigenfrequencies (often the lowest order mode) and thus make use of the high $Q$ by using the resonance phenomenon that will lead to large fields.

## 3.3 Accelerating voltage

We define the 'accelerating voltage' of a cavity (or more precisely of the considered cavity oscillation mode) as the integrated change of the kinetic energy of a traversing particle divided by its charge:

$$V_{acc} = \frac{1}{q}\int_{-\infty}^{\infty} q\left(\vec{E} + \vec{v}\times\vec{B}\right) ds, \tag{24}$$

where $ds$ denotes integration along the particle trajectory, taking the fields at the actual position of the particle at the time of passage. With the fields varying at a single frequency $\omega$ and particles moving with the speed $\beta c$ in the $z$ direction, this expression simplifies to

$$V_{acc} = \int_{-\infty}^{\infty} \underline{\vec{E}}(z) e^{j\frac{\omega}{\beta c}z}\, dz. \tag{25}$$

The underscore denotes now that we understand the field as the complex amplitude of the field of the cavity oscillation mode. The exponential accounts for the movement of the particle with speed $\beta c$ through the cavity while the fields continue to oscillate. It is clear that expression (25) is generally complex; the phase angle accounts for the phase difference between the RF field and the bunches of the passing beam; the complex amplitude is generally referred to as accelerating voltage.

Trying to use a homogeneous waveguide for acceleration, we can use the expression (13) as the dominating term for the field in Eq. (25). Using in addition Eq. (9), the integral to calculate the accelerating voltage will contain a term

$$\int e^{j\omega\left(\frac{1}{\beta c}-\frac{1}{v_{\phi,z}}\right)z}\, dz.$$

For this term not to vanish over large distances, the phase velocity of the mode $v_{\varphi,z}$ must be made equal to the speed of the particle beam, $\beta c$. Since, however, the phase velocity inside a waveguide is always larger than the speed of light, it is impossible to accelerate with a simple waveguide over large distances.

### 3.4 Transit-time factor

In our definition of the accelerating voltage, Eq. (25), we have accounted for the finite speed of the particles through the cavity. It thus includes already the so-called transit-time factor, which describes this effect alone. The transit-time factor $T$ can be understood as accelerating voltage normalized to a 'voltage'

$$\int_{-\infty}^{\infty} |\vec{E}(z)| \, dz,$$

and is defined as

$$T = \frac{|V_{acc}|}{\int_{-\infty}^{\infty} |\vec{E}(z)| \, dz}. \tag{26}$$

For the case of the simple pillbox cavity, in which the axial field is constant and in phase at every axial position, the integral (25) is easily calculated; the result is

$$T_{pillbox} = \frac{\left|\sin\left(\frac{\chi_{01} h}{2a}\right)\right|}{\frac{\chi_{01} h}{2a}} \tag{27}$$

and is plotted in Fig. 12. The transit time is close to 1 for small gaps, and its first zero occurs if the particle's transit time is equal to the RF period.

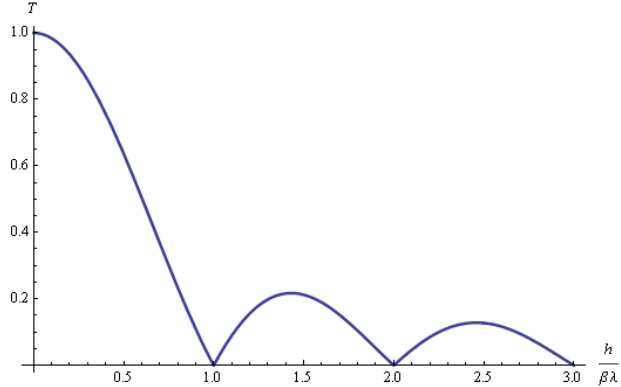

**Fig. 12:** Transit-time factor for a simple pillbox cavity

### 3.5 *R-upon-Q*, shunt impedance

Since the energy $W$ stored in the cavity is proportional to the square of the field (and thus the square of the accelerating voltage), it can be used to conveniently normalize the accelerating voltage; this leads to the definition of the quantity *R-upon-Q*:

$$\frac{R}{Q} = \frac{|V_{acc}|^2}{2\, \omega_0 W}. \tag{28}$$

The *R*-upon-*Q* thus simply quantifies how effectively the cavity converts stored energy into acceleration. Note that *R*-upon-*Q* is uniquely determined by the geometry of the cavity and not by the loss mechanism that leads to a finite *Q*.

Multiplying the *R*-upon-*Q* with the quality factor *Q*, one obtains the **shunt impedance** *R*, which describes how effectively the cavity converts input power into acceleration:

$$R = \left(\frac{R}{Q}\right) Q = \frac{|V_{acc}|^2}{2P}.  \quad (29)$$

Following this line of thought, the *R*-upon-*Q* may be considered a fundamental quantity and the shunt impedance *R* a derived quantity, in spite of the names that suggest otherwise. Note that there are a number of different definitions for these quantities in the technical literature. The definition given here ('circuit-ohms') is often used for synchrotrons, while in the definition used for linacs, the factor 2 on the right-hand side of Eqs. (28) and (29) is missing ('Linac-ohms').

It is interesting to see how much can be gained when trying to maximize the shunt impedance by just optimizing the shape of a cavity for given material and frequency. Figure 13 shows the shape and the electric field lines of the KEK 500 MHz photon factory cavity, which probably can be taken as a reference for this type of optimization. The inset indicates the values that could be obtained in this optimization, comparing it to the pillbox cavity with a height chosen to give the maximum shunt impedance. As can be seen from the indicated numbers, the gain is relatively modest — orders of magnitude can certainly not be reached.

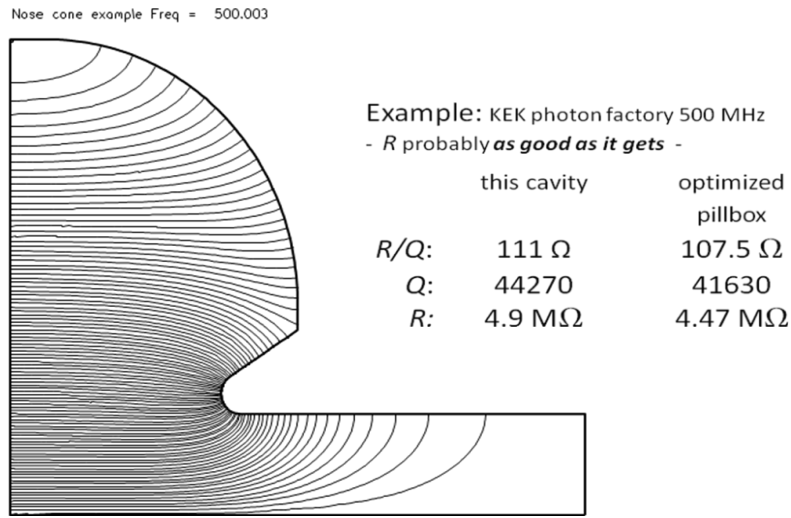

**Fig. 13:** Longitudinal section through half of a KEK photon-factory, 500 MHz, nose-cone cavity

## 3.6 Equivalent circuit

A cavity oscillation mode is conveniently described in an equivalent circuit as depicted in Fig. 14; driven by a current from the power source (or by the beam), the accelerating voltage develops across a parallel resonance circuit with resonance frequency $\omega_0$ and quality factor *Q*. Losses appear in its resistive element *R*; the name 'shunt impedance' now becomes obvious — it is 'shunting' the gap. The circuit equation

$$Y(\omega) \cdot V_{acc} = \frac{1}{R}\left(1 + jQ\left(\frac{\omega}{\omega_0} - \frac{\omega_0}{\omega}\right)\right) \cdot V_{acc} = I_G + I_B \quad (30)$$

describes what accelerating voltage can be obtained with which generator current. When setting the right-hand side of Eq. (30) equal to zero, we obtain an eigenvalue problem for $V_{acc}$ — the eigenvalues are the zeros of $Y(\omega)$ (or the poles of its inverse). The term proportional to $Q$ is responsible for the resonant behaviour of the cavity.

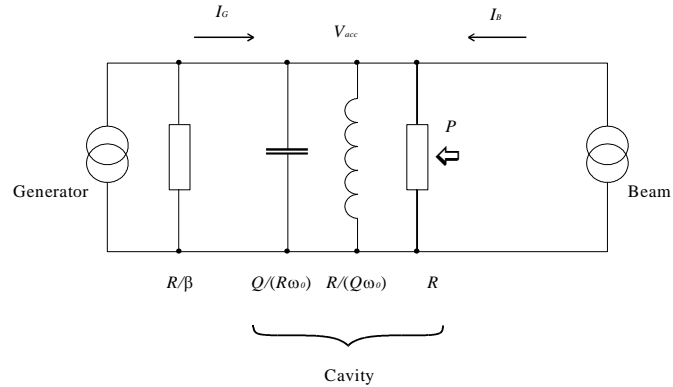

**Fig. 14:** Equivalent circuit of a resonant mode of a cavity

When plotting the accelerating voltage versus frequency for different values of $Q$ (Fig. 15), the resonance phenomenon becomes apparent. It is this phenomenon that allows developing large voltages with modest powers. Consequently one trend in RF technology development has been to optimize the $Q$ of cavities by design. Superconducting (SC) RF cavities are pushing this trend to the extreme; $Q$'s in the order of $10^{10}$ are typical for SC cavities. Also normal-conducting cavities use high $Q$'s to minimize the power losses; the technically obtained values depend on frequency and size and are typically in the range of some $10^4$.

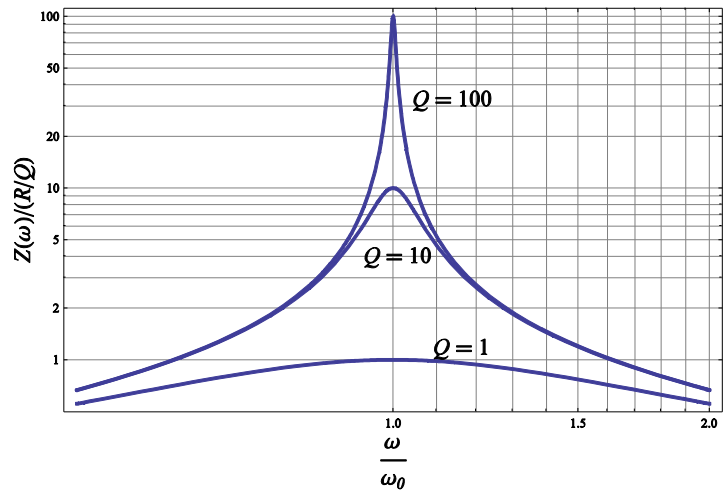

**Fig. 15:** Resonance behaviour of a cavity

But high $Q$'s also have disadvantages. As can be seen in Fig. 15, a high $Q$ leads to a very sharp resonance or a very narrow bandwidth resonator, which has to be tuned very precisely and may become very delicate and sensitive to error (machining tolerances, temperature, pressure, vibrations, etc.). A large stored energy will not allow for rapid changes of the field amplitude, frequency or phase. For a small ion synchrotron, for example, one may wish to apply RF with non-sinusoidal form and/or with rapidly varying frequency to the beam; these requirements call for cavities which either have a large instantaneous bandwidth (i.e., a low $Q$) or cavities that can be rapidly tuned. In these cases moderate or even extremely small $Q$'s can become optimum.

In addition to the fundamental mode, the field distribution of which is normally optimized for acceleration, other modes exist in the cavity, which can (and will) also interact with the particle beam. Even if not actively driven by an amplifier, these so-called higher order modes (HOMs) still present their impedance to the beam and may lead to instabilities and consequently have to be considered in the design. They are normally selectively coupled out and damped using external loads (HOM damper), thus reducing their *Q*.

## 4 Multi-gap cavities

### 4.1 How many gaps?

As we could see in Fig. 13 above, the shunt impedance of a normal-conducting cavity cannot be significantly increased. A few MΩ can be obtained after careful optimization, the exact value will depend on the frequency range. Limited by the available RF power and the cavity, this sets an upper limit to the accelerating voltage; for larger voltages one has to increase the number of RF systems and the power accordingly.

To introduce multi-gap cavities let us see what happens if one just increases the number of gaps, keeping the total power constant: consider *n* single-gap cavities with a shunt impedance *R*, as sketched in Fig. 16. The available power is split in equal parts and evenly distributed to the *n* cavities. According to Eq. (29), each cavity will produce an accelerating voltage of $\sqrt{2R(P/n)}$, so with the correct phasing of the RF, the total voltage will be just the sum, $V_{acc} = \sqrt{2(nR)P}$. If we now consider the assembly consisting of the *n* original cavities and the power splitter as a single cavity with *n* gaps, we notice that this new cavity has the shunt impedance *nR*; this is a significant increase. Consequently, by just multiplying the number of gaps, one can make much more efficient use of the available RF power to generate very large accelerating voltages.

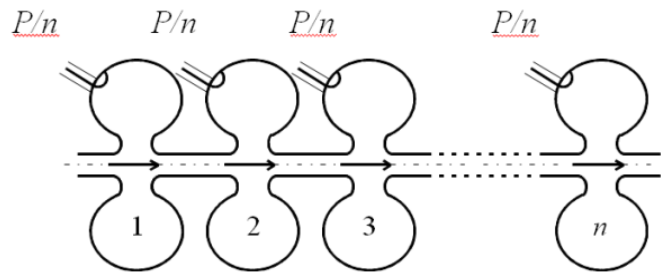

**Fig. 16:** Distributing a given power *P* to *n* cavities

Instead of using *n* individual power couplers and a large power splitter, much more elegant ways of distributing the available RF power to many gaps have been invented — one can combine the individual gaps in one vacuum vessel and one can in fact use this vacuum vessel itself as a distributed power splitter, which leads to standing-wave or travelling-wave cavities.

### 4.2 Travelling-wave structures

In a travelling wave structure, the RF power is fed via a power coupler into one end of the cavity (see Fig. 17), flowing (travelling) through the cavity, typically in the same direction as the beam, creating an accelerating voltage at every gap. An output coupler at the far end of the cavity is connected to a matched power load. If no beam is present, the input power reduced by the cavity losses goes to the power load where it is dissipated. In the presence of a large beam current, however, a large fraction of the forward-travelling power can be transferred to the beam, such that a much smaller power comes through the output coupler.

Different from the homogeneous waveguide, where the phase velocity is always larger than the speed of light, the phase velocity inside a travelling-wave structure can by design be made equal to the speed of the particles at the operating frequency, i.e., the phase advance over a cell of length *d* is equal to $2\pi d/\lambda$ for particles travelling with the speed of light; this allows acceleration over large distances. This is possible since travelling-wave structures are not homogeneous in the axial direction, but are periodic or almost periodic structures. Methods to analyse periodic structures have been developed for the analysis of periodic crystal lattices, looking at the wave function of electrons in a semiconductor for example. The result of such an analysis is the Brillouin diagram, which gives the 'modes' of the periodic structures; it shows the frequencies for given phase advances per cell. Like the frequency range below the lowest cutoff of a homogeneous waveguide, one finds frequency bands in which modes cannot propagate; for periodic structures they are referred to as stop-bands. In the pass-bands, where modes can propagate, the phase advance per cell varies from 0 to π. One can read the phase velocity of the mode directly in the diagram since it is simply the ratio $\omega/k$.. Figure 18 shows the Brillouin diagram for a travelling-wave structure. The straight line indicates the speed of light, the intersections with the Brillouin curve indicate combinations of frequency and phase advance per cell where the particle beam is synchronous to the wave — the structure is designed to operate at one of these intersections. If this condition is satisfied, one can imagine the particles 'surfing' on this forward-travelling wave.

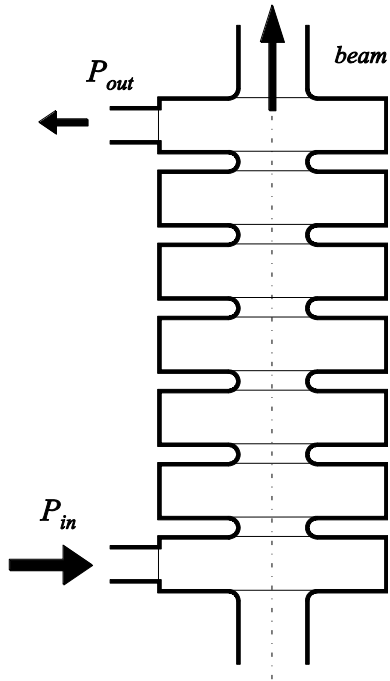 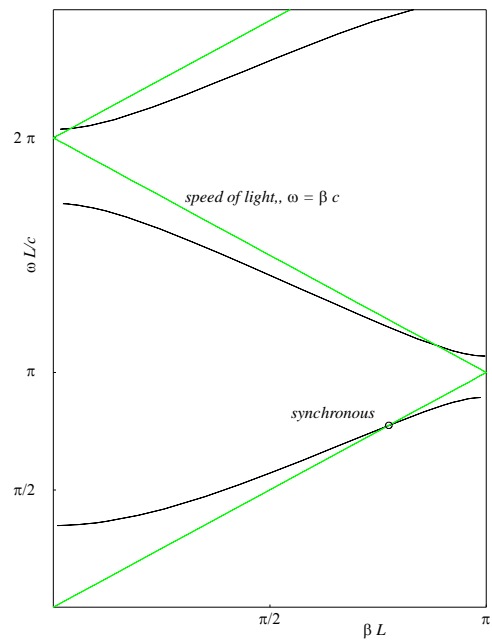

**Fig. 17:** Sketch of a travelling-wave structure     **Fig. 18:** A Brillouin diagram for a periodic structure